\documentclass[usenatbib]{mn2e}

\newif\ifAMStwofonts

\title[Detection of superimposed periodic signals]{Detection of superimposed periodic signals using wavelets}
\author[X. Otazu et al.]
       {X. Otazu$^{1,2}$\thanks{xotazu@am.ub.es}, M. Rib\'o$^1$, M. Peracaula$^1$, J. M. Paredes$^1$, J. N\'u\~nez$^{1,3}$ \\
        $^1$Departament d'Astronomia i Meteorologia, Universitat de Barcelona, Av. Diagonal 647, E-08028, Barcelona, Spain \\
$^2$Centre de Visi\'o per Computador, Edifici O, Campus Universitat Aut\`onoma de Barcelona (UAB), 08193 Cerdanyola (Barcelona), Spain \\
$^3$Observatori Fabra, Cam\'{\i} del Observatori sn, 08035 Barcelona, Spain}
\date{Accepted YYYY MMMMMMMM DD.
      Received YYYY MMMMMMMM DD}

\pagerange{\pageref{firstpage}--\pageref{lastpage}}
\pubyear{YYYY}

\begin{document}

\maketitle

\label{firstpage}

\begin{abstract}
In this paper we present a wavelet based algorithm that is able to detect
superimposed periodic signals in data with low signal-noise ratio. In this
context, the results given by classical period determination methods highly
depend on the intrinsic characteristics of each periodic signal, like amplitude
or profile. It is then difficult to detect the different periods present in the
data set. The results given by the wavelet based method for period
determination we present here are independent of the characteristics of the
signals.
\end{abstract}

\begin{keywords}
methods: data analysis -- stars: variables: other -- X-rays: general.
\end{keywords}

\section{Introduction}
\label{intro}

The search for periodic signals is common in many areas of astronomy, and, 
often, simple Fourier based methods or epoch folding methods are able to detect
the periods that are present in the data sets. However, this turns out to be
difficult when the periodic signals are not sinusoidal or when the 
signal-noise ratio is very low.

Specifically, in high energy astrophysics two problems are responsible for
reduced signal-noise ratios in the data: the relatively small number of high
energy photons when compared to the number of photons available at other
wavelengths, and the low efficiency in photon counting of high energy
detectors. Hence, the astronomical data supplied by X-ray telescopes,
especially flux monitorings, often suffer from poor statistics. This poses many
problems when processing and analysing the data in order to determine the flux,
detect periodic signals, etc. In addition, some X-ray sources present several
periods, which arise from different physical processes such as pulses, eclipses
in two-body systems or occultation by a precessing accretion disc. Some
examples of this kind of sources are SMC~X-1 \citep{wojdowski}, LMC~X-4
\citep{levine}, and LS~I~+65$^\circ$010 \citep{corbet}.

Classical period determination methods are based either on epoch folding
techniques or on Fourier decomposition analysis. The first group of methods are
based on the analysis of the dispersion of the diferent light curves produced
by folding the data over a range of trial periods (see, for example,
\citealt{lafler}; \citealt{jurkevich} and \citealt{stellingwerf}). The second
type of methods use the Fourier transform in combination with deconvolution
techniques to deal with the data sampling function (see, for example,
\citealt{lomb}; \citealt{scargleA}, 1989; \citealt*{roberts} and
\citealt{press}).

Epoch folding methods and Fourier based methods work well when applied to data
sets that present a unique period. However, when the analysed data set contains
several periodic signals, the behavior of these algorithms highly depends on
intrinsic signal characteristics. Every individual periodic signal may present
different spectral behavior (amplitude and profile) and depending on them, the
algorithm will detect some kinds of signal better than others. Fourier based
techniques will, in general, successfully detect two combined sinusoidal
signals, even if one of them has a low signal-noise ratio. However, they have
difficulties in detecting the non-sinusoidal signals that might be present in a
data set (as in the astronomical case of pulsed emission superimposed on an
orbital period). On the contrary, epoch folding methods are well suited for
detecting non-sinusoidal signals but they tend to fail when two or more periods
are present in the data set, especially when the signal-noise ratio is low.

In general, we can say that the greater the difference between the spectral
characteristics of  each signal, the more difficult it is to detect each period
with classical methods, and the less statistically significant are the results
given by them. Therefore, a new algorithm becomes immediately necessary for the
detection of superimposed periodic signals.

In this paper, we show how the methodology of wavelet theory is very well
suited to this problem, since it is completely oriented to decompose functions
into their several spectral characteristics. This allows us to isolate every
signal present in our data and to analyse them separately, avoiding their
mutual influences. In Section 2 we outline some concepts in wavelet theory
relevant to the stated problem. In Section 3 we propose an algorithm to detect
each of the periodic signals present in a data set by combining wavelet
decomposition with classical period determination methods. In Sections 4 and 5
we present some examples of synthetic data we used to test the algorithm and
the results we obtained. We summarise our conclusions in Section 6.

\section{Outline of the wavelet transform}
\label{outline}

Multiresolution analysis based on the wavelet theory introduces the concept of
details between successive levels of scale or resolutions (\citealt{chui};
\citealt{daubechies}; \citealt{meyer}; \citealt{young}; \citealt{kaiser}; 
\citealt{vetterli}).

Wavelet decomposition is increasingly being used for astronomical signal
processing (\citealt{starckA}; \citealt{starckB}) and remotely sensed data
(\citealt{yocky}; \citealt{datcu}). The method is based on the decomposition of
the signal into multiple channels based on their local frequency content. The
wavelet transform is an intermediate representation between the Fourier and the
temporal one. In the Fourier transform we know the global frequency content of
our signal, but we have no information about the temporal localization of these
frequencies. However, the wavelet transform gives us an idea of both the local
frequency content and the temporal distribution of these frequencies. Since in
Fourier space the basis functions are sinusoidal, they extend along all space
and do not have temporal concentration. However, wavelets are concentrated
around a central point, thus they have a high degree of temporal localization.

\subsection{Some theory}

Next, we will briefly present an outline of the wavelet transform, relying on
the multiresolution signal representation concept.

Given a signal $f(t)$, we construct a sequence $F_m(f(t))$ of approximations of
$f(t)$. Each $F_m(f(t))$ is specific for the representation of the signal at a
given scale (resolution). $F_m(f(t))$ represents the projection of the signal
$f(t)$ from the signal space $S$ onto subspace $S_m$. In this representation,
$F_m(f(t))$ is the `closest' approximation of $f(t)$ with resolution $2^m$.

The differences between two consecutive scales $m$ and $m+1$ are the
multiresolution wavelet planes or `detail' signal at resolution $2^m$:
\begin{equation}
w_m(f(t)) = F_m(f(t)) - F_{m+1}(f(t))\;.
\end{equation}
This detail signal can be expressed as:
\begin{equation}
w_m(f(t)) = \sum_l{W_{m,l}(f) \psi_{m,l}(t)}\;,
\end{equation}
where coefficients $W_{m,l}(f)$ are given by the {\it direct wavelet transform of the signal} f(t):
\begin{equation}
W_{m,l}(f) = \int_{-\infty}^{+\infty}
f(t) \,\psi_{m,l}(t)~dt\;.
\label{WT_cont}
\end{equation}

The coefficients $W_{m,l}(f)$ are called wavelet coefficients of $f(t)$. Such
coefficients correspond to the fluctuations of the signal $f(t)$ near the point
$l$ at resolution level $m$. Thus, the wavelet transform~(\ref{WT_cont})
represents the expansion of signal $f(t)$ in the set of basis functions
$\psi_{m,l}(t)$. These basis functions are scaled and translated versions of a
general function $\psi(t)$ called Mother Wavelet. To construct the basis
functions $\psi_{m,l}(t)$, the Mother Wavelet is dilated and translated
according to parameters $m$ and $l$ as follows:
\begin{equation}
\psi_{m,l}(t) = 2^{m/2} \psi(2^mt - l)\;.
\label{basis}
\end{equation}
Therefore, all the basis functions $\psi_{m,l}(t)$ have the same profile, that
is, the Mother Wavelet profile. Using~(\ref{basis}) we obtain an orthonormal
wavelet basis \citep{daubechies}.

The {\it inverse discrete wavelet transform} is given by the reconstruction
formula:
\begin{equation}
f(t) = \sum_m {\sum_l{W_{m,l}(f) \psi_{m,l}(t)}}\;.
\end{equation}

In summary, the wavelet transform describes at each resolution step the
difference between the previous and the current resolution representation. By
iterating the process from the highest to the lowest available resolution level
we obtain a pyramidal representation of the signal. This usually includes a
decimation process, i.e., in each iteration 1 out of 2 points is discarded,
which implies that the number of data points at lower frequencies is highly
reduced.

In the Fourier transform, the noise contribution is spread along all
frequencies. In contrast, one of the interesting properties of the wavelet
transform (which is also frequency based) is that noise contribution is more
important on the higher frequency wavelet planes.

\subsection{The `\`{a} trous' algorithm}

In order to obtain a discrete wavelet decomposition for signals, we follow
\cite{starckA} and we use the discrete wavelet transform algorithm known as
`\`{a} trous' (`with holes') to decompose the signal into wavelet planes. Given
a signal $p$ we construct the sequence of approximations:
\begin{equation}
F_1(p) = p_1,\;\;
F_2(p_1) = p_2,\;\; 
F_3(p_2) = p_3,\cdots\;.
\end{equation}

To construct the sequence, this algorithm performs successive convolutions with
a filter obtained from an auxiliary function named scaling function. We use a
scaling function which has a $B_3$ cubic spline profile. The use of a $B_3$
cubic spline leads to a convolution with a mask of 5 elements, all elements
being scaled up by 16: (1,4,6,4,1).

As stated above, the wavelet planes are computed as the differences between two
consecutive approximations $p_{i-1}$ and $p_i$. Letting $w_i = p_{i-1} - p_i
\;\;(i=1,\cdots,n)$, in which $p_0 = p$, we can write the reconstruction
formula:
\begin{equation}
p = \sum_{i=1}^{n}w_i + p_r\;\;.
\label{wav_des}
\end{equation}

In this representation, the signals $p_i~(i=0,\cdots,n)$ are versions of the
original signal $p$ at increasing scales (decreasing resolution levels),
$w_i~(i=1,\cdots,n)$ are the multiresolution wavelet planes and $p_r$ is a
residual signal (in fact $n=r$, but we explicitly substitute $n$ by $r$ to
clearly express the concept of $residual$). In our case, we are using a dyadic
decomposition scheme. Thus, the original signal $p_0$ has double resolution
than $p_1$, the signal $p_1$ double resolution compared to $p_2$, and so on. If
the resolution of signal $p_0$ is, for example, $10\,\Delta t$ (being $\Delta
t$ the sampling of the data), the resolution of $p_1$ would be $20\,\Delta t$,
the resolution of $p_2$ would be $40\,\Delta t$, etc. All these
$p_i~(i=0,\cdots,n)$ signals have the same number of data points, in contrast
to some more usual wavelet decomposition algorithms which reduce the number of
points by a factor 2 when going through increasing scales (process known as
decimation step). Later we will see why we are interested in maintaining the
number of points.

\section{Proposed algorithm}
\label{algorithm}

We propose to apply the wavelet decomposition using the `\`{a} trous' algorithm
to solve our initial stated problem: to isolate each of the periodic signals
contained in a set of data and study them separately.

Hence, the algorithm we propose is as follows:

\begin{enumerate}

\item Choose a value for $n$ and decompose the original signal $p$ into its
wavelet planes $w_i~(i=0,\cdots,n)$.

\item Detect periods in each of the $n$ obtained wavelet plane $w_i$.

\end{enumerate}

Any method can be used to detect the periods present in each $w_i$. In our
tests, we used Phase Dispersion Minimization ({\sevensize PDM})
\citep{stellingwerf} and {\sevensize CLEAN} \citep{roberts} methods.
{\sevensize PDM} belongs to the group of epoch folding methods mentioned in the
introduction. Therefore it folds the data points over a set of trial periods
and analyses the dispersion of every light curve in the phase space. To do
that, the data points are grouped in phase bins and the most probable period is
then found by minimizing the sum of the bin variances. On the other hand,
{\sevensize CLEAN} belongs to the group of Fourier transform decomposition
methods. It works in the frequency domain and it is based on the fact that the
spectrum obtained from the Fourier transformation of the data set contains
artefacts caused by the imcompleteness of the sampling (time gaps) and the
finite time span of it. CLEAN uses a non-linear deconvolution algorithm to
clean (hence its name) these artefacts from the original (dirty) spectrum. 
This is done by interatively substracting from the dirty spectrum the expected
response of a signal composed by a unique harmonical function (clean
component). From the set of clean components, a clean spectrum is obtained
which is at the end corrected to preserve the noise level and the frequency
resolution.

{\sevensize PDM} is well suited to identify signals with any profile, not just
a sinusoidal one. For instance, if we have a burst-like periodic signal, it is
more successfully detected by the {\sevensize PDM} algorithm than by
{\sevensize CLEAN}. On the other hand, if we knew that we are trying to
estimate the period of a sinusoidal signal, it would be better to use
{\sevensize CLEAN}.

The algorithm we present here could be seen as an improvement over the usual
period detection algorithms, because prior to using them we decompose our
signal into its $w_i$ wavelet planes, and afterwards we apply these algorithms
to each $w_i$.

In {\sevensize PDM} and {\sevensize CLEAN} methods there is only one data set
to study and to estimate its period. But in the wavelet based method there are
several data sets to study (several wavelet planes), and this improves the
detection probabilities. We can search for a certain period in several wavelet
planes, which helps to discard some of the usual marginal artefact detections
present in {\sevensize PDM} and {\sevensize CLEAN} and to detect the true
periods present in our original data.

We can use other wavelet based algorithms \citep{szatmary}, which are based on
approximations of a continuous wavelet transform and the study of wavelet space
coefficients. However, these algorithms present a {\it non direct} inverse
wavelet transform: the search for periodicities is based on the fit between the
wavelet base function profile and the signal one, and therefore on the values
of the wavelet transform coefficients, which highly depend on the wavelet base
used. For example, if we took a signal constructed by a Gaussian with 1 day
duration which repeats every 30 days, it would be very difficult to find a
wavelet base that fits well to this sparse signal. On the contrary, if we take
a suitable wavelet base and a decomposition scheme which allows us to perform
easily its inverse transform, this base will fit the profile of every Gaussian.
Hence, performing the inverse transform of every wavelet scale, we can work in
the temporal representation to find directly the periodicities. This is the
case of the wavelet decomposition algorithm we use in the method we present
here.

Another reason to use the `\`{a} trous' algorithm is that the sampling of the
wavelet planes is the same as the original data. This allows us to work
directly in the temporal space with the frequency content of the corresponding
wavelet plane. In other wavelet decomposition algorithms, we are usually forced
to work on the decimated wavelet space.

Hereafter, and for notational convenience, the wavelet based {\sevensize PDM}
and {\sevensize CLEAN} methods will be called {\sevensize WPDM} and {\sevensize
WCLEAN} respectively.

\section {Simulated data}
\label{simulated}

In order to check the benefit of applying {\sevensize WPDM} versus {\sevensize
PDM}, or {\sevensize WCLEAN} versus {\sevensize CLEAN}, we generated several
sets of simulated data containing two superimposed periodic signals. Each data
set is composed of a high amplitude primary sinusoidal function and a secondary
low amplitude one. For the latter we used two kinds of functions: sinusoidal
and Gaussian. The first are intended to simulate variable sources with pure
sinusoidal intensity profile (like precession of accretion discs), and the
second burst-like events (like pulses or eclipses). Finally, we added a
white-Gaussian noise to this combination of signals. We increased the value of
the noise standard deviation, $\sigma$, up to the value where detection of
periodic signals became statistically insignificant in both the classical and
the wavelet based methods.

Here we list the characteristics of the signals:

\begin{enumerate}

\item Each signal is generated as an evenly spaced data set composed of 1000
points, separated by a unit between them.

\item The high amplitude sinusoidal function has an amplitude equal to 1 and a
period of 108.5 units. The period selected should satisfy two conditions: not
be an exact divisor of the number of points in the data set, and allow the
presence of more than 5 complete periods in the simulated signal to guarantee
the possibility of detection.

\item The amplitudes of the low amplitude periodic functions (sinusoidal or
Gaussian) are 0.1 and 0.5.

\item The periods used for the secondary functions are 13.13 and 23.11 units.
Others could be used, satisfying the condition that they are not harmonics of
the high amplitude sinusoidal function period.

\item In the case of the Gaussian signal, we have used two values for the Full
Width Half Maximum (FWHM), corresponding to 2 and 6 units, respectively.

\end{enumerate}

In the first three columns of Table~\ref{sine_table} we present the parameters
used to generate each simulated data set in the case of sine+sine periodic
signals. In Table~\ref{gauss_table} the first four columns show the parameters
used to generate the sine+Gaussian signals.

\section{Results}
\label{results}

The simultaneous use of two independent methods, such as {\sevensize CLEAN} and
{\sevensize PDM}, is usually applied to discriminate false period detections
from the true ones. A similar procedure can be used with each one of the
wavelet planes in the {\sevensize WPDM} and {\sevensize WCLEAN} methods.
Therefore, when comparing the behavior of these different methods, we have to
compare the usual {\sevensize PDM-CLEAN} method combination for period
estimation prior to the new {\sevensize WPDM-WCLEAN} combination.

The primary period (108.5 units) is always detected by all methods, and does
not appear in Table~\ref{sine_table} and Table~\ref{gauss_table}. In these
tables, the four last columns show the detected low amplitude periods for each
data set using {\sevensize PDM}, {\sevensize CLEAN}, {\sevensize WPDM} and
{\sevensize WCLEAN} methods, respectively. A dash is shown when a period is not
detected, and a question mark when the detection is difficult or doubtful. When
a period is indicated in the wavelet based methods, we also show in parentheses
the wavelet planes where it is detected.

\begin{figure}
\mbox{}
\vspace{6.7cm}
\includegraphics{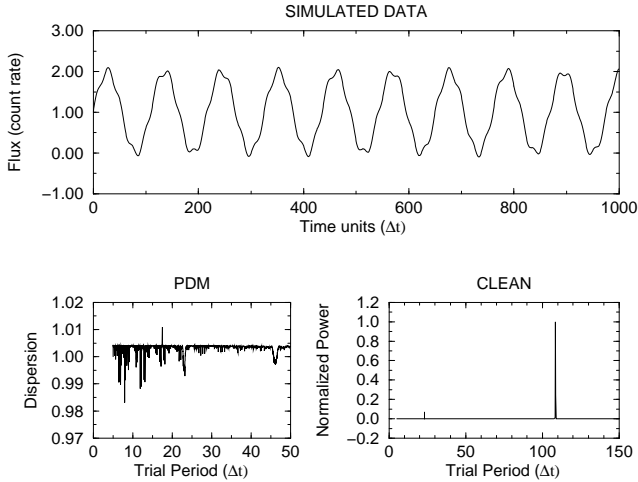}
\caption[]{Top: simulated data of a sinusoid with 23.11 units period, amplitude=0.1 and $\sigma_{\mathrm noise}=0$, superimposed on the sinusoid with 108.5 units period and amplitude=1.0. Bottom: {\sevensize PDM} and {\sevensize CLEAN} outputs of this data set.}
\label{sine_23_1}
\end{figure}

\subsection{Sine+sine case}

\begin{figure}
\mbox{}
\vspace{6.0cm}
\includegraphics{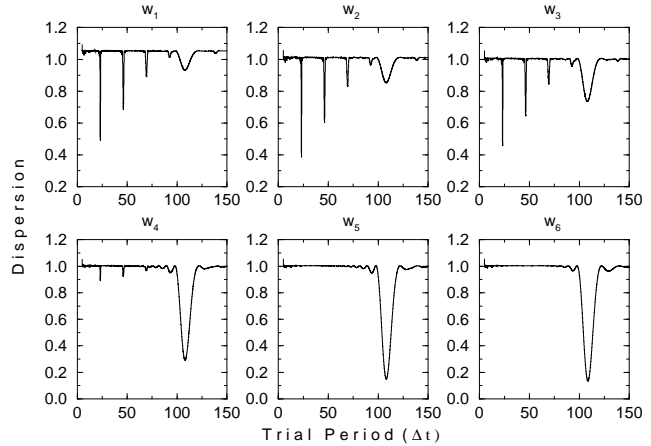}
\caption[]{{\sevensize WPDM} output for each wavelet plane of the data at the top of Fig.~\ref{sine_23_1}.}
\label{sine_23_1.wpdm}
\end{figure}

\begin{figure}
\mbox{}
\vspace{6.0cm}
\includegraphics{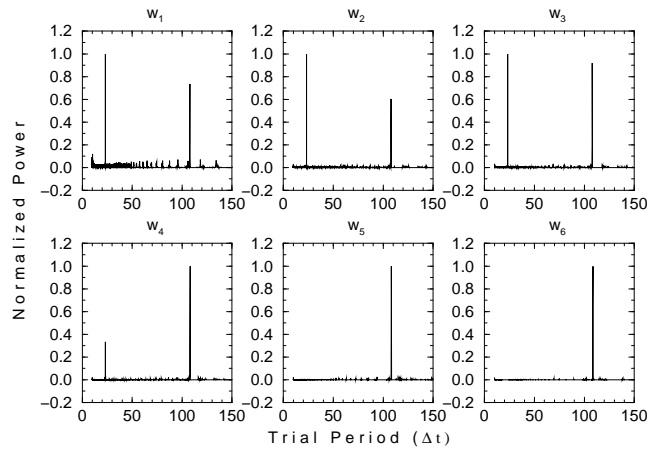}
\caption[]{{\sevensize WCLEAN} output for each wavelet plane of the data at the top of Fig.~\ref{sine_23_1}.}
\label{sine_23_1.wclean}
\end{figure}

To illustrate the performances of {\sevensize WPDM} and {\sevensize WCLEAN}
methods, we show in Fig.~\ref{sine_23_1} a simulated data set which contains
two pure sinusoidal functions without noise, and both their {\sevensize PDM}
and {\sevensize CLEAN} outputs (in {\sevensize PDM} we look for minima with
subharmonics, while in {\sevensize CLEAN} we look for maxima). The 108.5 units
period is detected by both methods. In the case of {\sevensize PDM} we have
zoomed into the range of short trial periods and dispersion to try to detect
the 23.11 units minimum. It is clear that even in the case of signals without
noise, {\sevensize PDM} method is unable to find the secondary signal
(period=23.11 units). In contrast, {\sevensize CLEAN} is able to find it,
although with a low normalized power (less than 0.1, which is the ratio between
the amplitudes of both periodic signals). The output of {\sevensize WPDM} is
shown in Fig.~\ref{sine_23_1.wpdm} and the one of {\sevensize WCLEAN} in
Fig.~\ref{sine_23_1.wclean}. As one would expect, the secondary period is only
seen in the lower wavelet planes ($w_1$ to $w_4$), while the primary long-term
period of 108.5 units is better detected when working with higher wavelet
planes ($w_4$ to $w_6$). We have worked until $w_6$, because in this plane the
108.5 units period is clearly detected, and there is no reason to continue with
higher values of $n$. This will be the case in all the analysed data we present
here.

\begin{table*}
 \centering
 \begin{minipage}{116mm}
\caption[]{\label{sine_table} Periods detected in the sine+sine data sets. The first three columns are the parameters used for the simulated data. A dash is shown when a period is not detected, and a question mark when the detection is difficult or doubtful. When a period is indicated in the wavelet based methods, we also show in parentheses the wavelet planes where it is detected.}
\begin{tabular}{cccllll}
\hline
Period & Amplitude & $\sigma_{\mathrm noise}$ & {\sevensize PDM} & {\sevensize CLEAN} & {\sevensize WPDM} & {\sevensize WCLEAN} \\
\hline
13.13 & 0.1& 0.0   &  $-$    & 13.13  & 13.13 (1,2,3)   & 13.13 (1,2,3) \\
  &        & 0.1   &  $-$  & 13.12  & 13.13 (2,3)     & 13.13 (2,3) \\
  &        & 0.2   &  $-$  & 13.11  & 13.11 (2,3)     & 13.11 (2,3) \\
  &        & 0.4   &  $-$  & 13.09  & 13.09 (2,3)     & 13.09 (2,3) \\
  &        & 0.6   &  $-$  & 13.09? & 13.09 (3)       & 13.09 (2,3) \\
  &        & 0.8   &  $-$  & $-$    & 13.08 (3)       & 13.08 (2,3) \\
  &        & 1.0   &  $-$  & $-$    & 13.08 (3)       & 13.08 (2,3) \\
  &        & 1.2   &  $-$  & $-$    & 13.07 (3?)      & 13.07 (2?,3?) \\
  &        & 1.4   &  $-$  & $-$    & 13.07 (3?)      & 13.07 (2?,3?) \\
  &        & 1.6   &  $-$  & $-$    & $-$             & $-$ \\
  & 0.5    & 0.0   & 13.13 & 13.13  & 13.13 (1,2,3)   & 13.13 (1,2,3) \\
  &        & 0.5   & 13.12 & 13.12  & 13.13 (2,3)     & 13.13 (2,3) \\
  &        & 1.0   & 13.12 & 13.12  & 13.13 (2,3)     & 13.13 (2,3) \\
  &        & 2.0   & 13.09 & 13.09  & 13.13 (2,3)     & 13.13 (2,3) \\
  &        & 3.0   & 13.08?& 13.08? & 13.13 (2?,3)    & 13.13 (2,3) \\
  &        & 4.0   & $-$   & $-$    & 13.13 (3)       & 13.13 (2,3) \\
  &        & 5.0   & $-$   & $-$    & 13.13 (3)       & 13.13 (2?,3) \\
  &        & 6.0   & $-$   & $-$    & 13.13 (3?)      & 13.13 (3?) \\
  &        & 8.0   & $-$   & $-$    & $-$             & $-$ \\
23.11 & 0.1& 0.0   & $-$   & 23.11  & 23.11 (1,2,3,4) & 23.11 (1,2,3,4) \\
  &        & 0.1   & $-$   & 23.12  & 23.11 (2,3,4)   & 23.11 (2,3,4) \\
  &        & 0.2   & $-$   & 23.12  & 23.13 (3,4)     & 23.13 (3,4) \\
  &        & 0.4   & $-$   & 23.12? & 23.11 (3,4)     & 23.11 (3,4) \\
  &        & 0.6   & $-$   & $-$    & 23.17 (3,4)     & 23.17 (3,4) \\
  &        & 0.8   & $-$   & $-$    & 23.13 (3?,4)    & 23.13 (3?,4) \\
  &        & 1.0   & $-$   & $-$    & 23.13 (4?)      & 23.11 (4?) \\
  &        & 1.2   & $-$   & $-$    & $-$             & $-$ \\
  & 0.5    & 0.0   & 23.11 & 23.11  & 23.11 (1,2,3,4) & 23.11 (1,2,3,4) \\
  &        & 0.5   & 23.11 & 23.11  & 23.11 (2,3,4)   & 23.11 (1,2,3,4) \\
  &        & 1.0   & 23.10 & 23.12  & 23.12 (2?,3,4)  & 23.13 (2?,3,4) \\
  &        & 2.0   & 23.11 & 23.11  & 23.11 (3,4)     & 23.11 (3,4) \\
  &        & 3.0   & $-$   & 23.11? & 23.17 (3?,4)    & 23.17 (3?,4) \\
  &        & 4.0   & $-$   & $-$    & 23.13 (4?)      & 23.13 (4?) \\
  &        & 5.0   & $-$   & $-$    & $-$             & $-$ \\
\hline
\end{tabular}
\end{minipage}
\end{table*}

Some remarks can be made after inspection of Table~\ref{sine_table}:

\begin{enumerate} 

\item There is a better performance of {\sevensize CLEAN} relative to
{\sevensize PDM}, because the latter has problems when dealing with
superimposed low amplitude periodic signals. Only when the amplitude of the
secondary signal is close enough to the primary one (amplitude=0.5) is
{\sevensize PDM} able to detect it.

\item In all methods, with high noise-signal ratios the detected periods are
slightly different from the simulated ones.

\item {\sevensize WPDM} and {\sevensize WCLEAN} perform always better than
{\sevensize PDM} and {\sevensize CLEAN} methods. When {\sevensize CLEAN}
marginally detects the secondary period, {\sevensize WPDM} and {\sevensize
WCLEAN} have no problems to detect it, and they work properly even with higher
noise. In the {\sevensize WPDM} case, the results are always much better than
with {\sevensize PDM}.

\item As the noise increases, the detection starts to fail in the lower wavelet
planes (higher frequencies), and only the higher ones (lower frequencies) are
noise-free enough to allow period detection.

\item The maximum ratio between the noise and the signal amplitude to detect a
period with {\sevensize WPDM} and {\sevensize WCLEAN}, is nearly constant for a
given period, being around 12 for the 13.13 units period and around 8 for the
23.11 units period. The reason for this difference is that there are 76
complete periods in the case of 13.13 units period, and only 43 complete
periods in the case of 23.11 units period.

\end{enumerate}

\begin{figure}
\mbox{}
\vspace{6.0cm}
\includegraphics{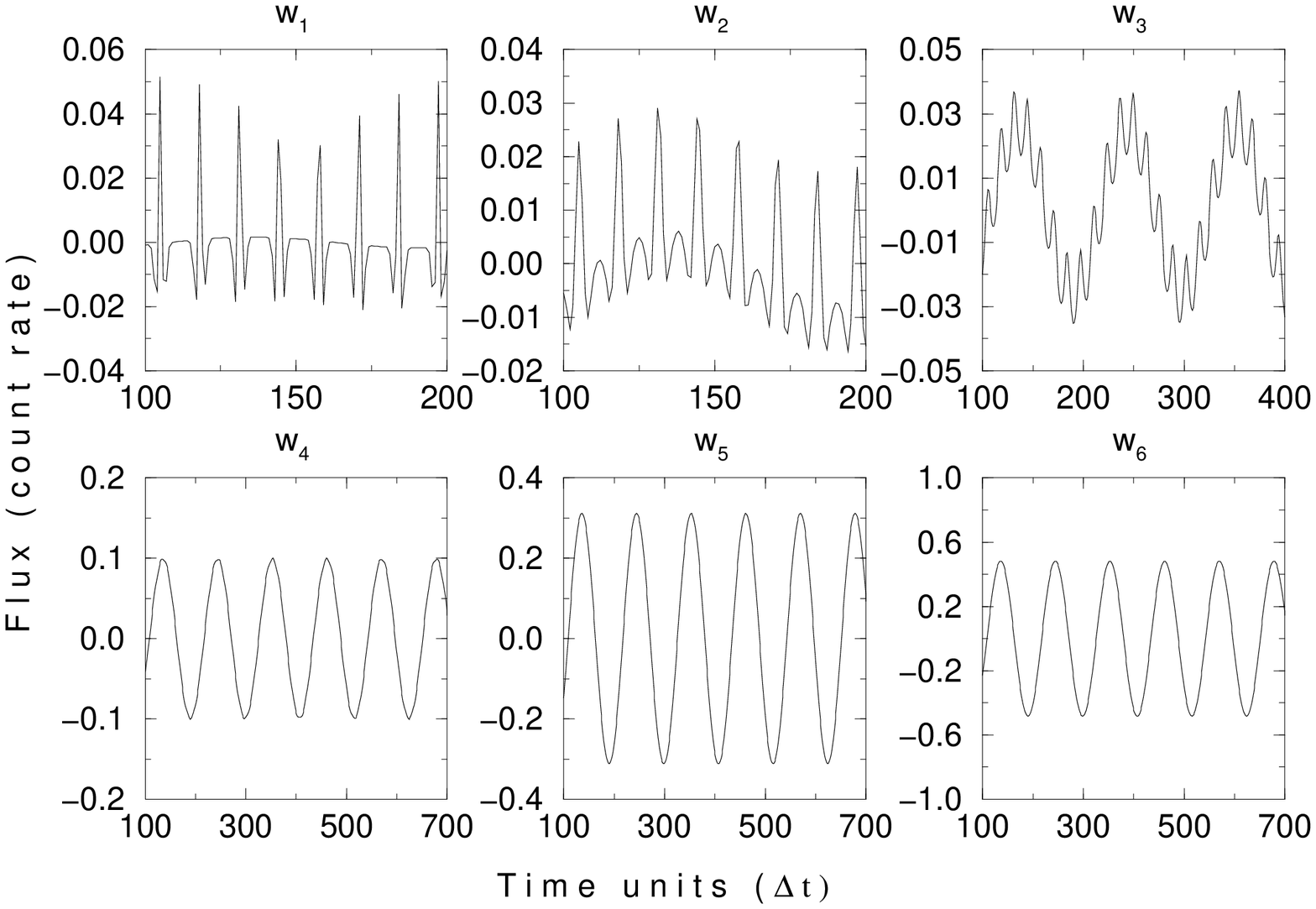}
\caption[]{Wavelet decomposition of the simulated data which contains a Gaussian with 13.13 units period, a FWHM of 2.0 units, amplitude=0.1 and $\sigma_{\mathrm noise}=0$, superimposed on the sinusoid with 108.5 units period and amplitude=1.0.}
\label{dwd_gauss_13_2_0.1}
\end{figure}

\begin{table*}
\centering 
 \begin{minipage}{133mm}
\caption[]{\label{gauss_table} Periods detected in the sine+Gaussian data sets. The first four columns are the parameters used for the simulated data. A dash is shown when a period is not detected, and a question mark when the detection is difficult or doubtful. When a period is indicated in the wavelet based methods, we also show in parentheses the wavelet planes where it is detected.}
\begin{tabular}{ccccllll}
\hline
FWHM & Period & Amplitude & $\sigma_{\mathrm noise}$ & {\sevensize PDM} & {\sevensize CLEAN} & {\sevensize WPDM} & {\sevensize WCLEAN} \\
\hline
2.0&13.13&0.1& 0.0 &  $-$  & 13.13  & 13.13 (1,2,3)    & 13.13 (1,2,3) \\
 &   &     & 0.1   &  $-$  & $-$    & 13.13 (2?,3?)    & 13.11 (3?) \\
 &   &     & ~0.15 &  $-$  & $-$    & $-$              & $-$ \\
 &   & 0.5 & 0.0   &  $-$  & 13.13  & 13.13 (1,2,3)    & 13.13 (1,2,3) \\
 &   &     & 0.5   &  $-$  & 13.14  & 13.14 (2,3)      & 13.14 (2?,3) \\
 &   &     & ~0.75 &  $-$  & $-$    & 13.15 (2?,3?)    & $-$ \\
 &   &     & 1.0   &  $-$  & $-$    & $-$              & $-$ \\
 & 23.11 & 0.1 & 0.0   &  $-$  & 23.11  & 23.11 (2?,3)     & 23.11 (3?,4) \\
 &   &   & 0.1         &  $-$  & $-$    & 23.11 (2?,3?)    & 23.11 (4?) \\
 &   &   & ~0.15       &  $-$  & $-$    & $-$              & $-$ \\
 &   & 0.5 & 0.0    &  $-$  & 23.11  & 23.11 (2?,3,4)   & 23.11 (2?,3) \\
 &   &     & 0.5    &  $-$  & $-$    & 23.12 (2?,3?)    & $-$ \\
 &   &     & ~0.75  &  $-$  & $-$    & $-$              & $-$ \\
6.0&13.13&0.1 & 0.0 &  $-$   & 13.13  & 23.12 (1,2,3)    & 13.13 (2,3)\\
 &   &     & 0.1    &  $-$   & 13.13  & 23.13 (2,3)      & 13.13 (2,3)\\
 &   &     & 0.2    &  $-$   & 13.13? & 23.14 (2?,3)     & 13.13 (2?,3)\\
 &   &     & 0.3    &  $-$   & $-$    & 23.14 (3)        & 13.13 (2?,3) \\
 &   &     & 0.5    &  $-$   & $-$    & $-$              & $-$ \\
 &   & 0.5 & 0.0  & 13.13  & 13.13  & 13.13 (1,2,3,4?) & 13.13 (1,2,3,4?) \\
 &   &     & 0.5  & 13.13  & 13.13  & 13.13 (2,3)      & 13.13 (2,3) \\
 &   &     & 1.0  & 13.12? & 13.12? & 13.13 (2?,3)     & 13.13 (2?,3) \\
 &   &     & 1.5  & $-$    & $-$    & 13.13 (3)        & $-$ \\
 &   &     & 2.0  & $-$    & $-$    & $-$              & $-$ \\
 &23.11&0.1& 0.0   & $-$    & 23.11  & 23.11 (1,2,3,4?) & 13.13 (1,2,3,4?) \\
 &   &     & 0.1   & $-$    & 23.12  & 23.11 (2,3,4?)   & 23.11 (2,3,4?) \\
 &   &     & ~0.15 & $-$    & 23.12? & 23.12 (2?,3,4?)  & 23.11 (2?,3,4?) \\
 &   &     & 0.2   & $-$    & $-$    & $-$              & $-$ \\
 &   & 0.5 & 0.0  & 23.11  & 23.11  & 23.11 (1,2,3,4)  & 13.13 (1,2,3,4) \\
 &   &     & 0.5  & 23.11  & 23.11  & 23.11 (1,2,3)    & 23.11 (2,3,4) \\
 &   &     & 1.0  & 23.12? & 23.12? & 23.12 (3,4)      & 23.11 (3,4?) \\
 &   &     & 1.5  & $-$    & $-$    & 23.12 (3?)       & $-$ \\
 &   &     & 2.0  & $-$    & $-$    & $-$              & $-$ \\
\hline
\end{tabular}
\end{minipage}
\end{table*}

\subsection{Sine+Gaussian case}

For illustrative purposes, we show in Fig.~\ref{dwd_gauss_13_2_0.1} the wavelet
decomposition of a sine+Gaussian signal. In the lower wavelet planes (planes
$w_1$ and $w_2$) we have isolated the Gaussian contribution. In the $w_3$ we
have a mixed contribution from both periods, but in the higher planes we only
have the longest period contribution.

We must note that the use of two different FWHM for the Gaussian, combined with
two different periods (13.13 and 23.11 units), gives 4 different profiles.
Hence, the phase duration of the burst-like event ranges from very low to
relatively high values in the following order: FWHM=2 and period=23.11, FWHM=2
and period=13.13, FWHM=6 and period=23.11, and finally FWHM=6 and period=13.13.

In view of the results displayed in Table~\ref{gauss_table} we can make the
following comments:

\begin{enumerate}

\item Again, there is a better performance of {\sevensize CLEAN} over
{\sevensize PDM}, for the reason explained above. However, we must note that
when the FWHM is only 2 units, {\sevensize PDM} never detects the secondary
period. Only with FWHM=6 units and an amplitude of 0.5, can {\sevensize PDM}
detect the low amplitude periodic signals.

\item In all methods, with high noise-signal ratios the detected periods are
slightly different from the simulated ones.

\item {\sevensize WPDM} and {\sevensize WCLEAN} perform always better than
{\sevensize PDM} and {\sevensize CLEAN} methods. When {\sevensize CLEAN}
marginally detects the secondary period, {\sevensize WPDM} and {\sevensize
WCLEAN} have no problems to detect it, and they work properly even with higher
noise. In the {\sevensize WPDM} case, the results are always much better than
with {\sevensize PDM}.

\item As the noise increases, the detection starts to fail in the lower wavelet
planes (higher frequencies), and only the higher ones (lower frequencies) are
noise-free enough to allow period detection.

\item In all cases with amplitude=0.1, the {\sevensize WPDM} performance is
very similar to {\sevensize WCLEAN}. In the amplitude=0.5 cases, {\sevensize
WPDM} is always better than {\sevensize WCLEAN} because the signal is not
sinusoidal and the amplitude is high enough to allow a good detection.

\item For a given amplitude of the Gaussian signal, the maximum noise-signal
ratio achieved with {\sevensize WPDM} and {\sevensize WCLEAN} increases with
the phase duration of the FWHM.

\end{enumerate}

\begin{figure}
\mbox{}
\vspace{6.7cm}
\includegraphics{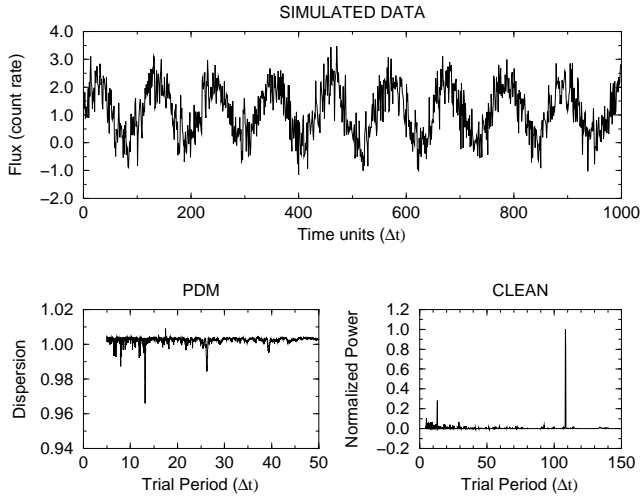}
\caption[]{Top: simulated data of a Gaussian with 13.13 units period, a FWHM of 6.0 units, amplitude=0.5 and $\sigma_{\mathrm noise}=0.5$, superimposed on the sinusoid with 108.5 units period and amplitude=1.0. Bottom: {\sevensize PDM} and {\sevensize CLEAN} outputs of this data set.}
\label{gauss_13_0.5}
\end{figure}

\begin{figure}
\mbox{}
\vspace{6.0cm}
\includegraphics{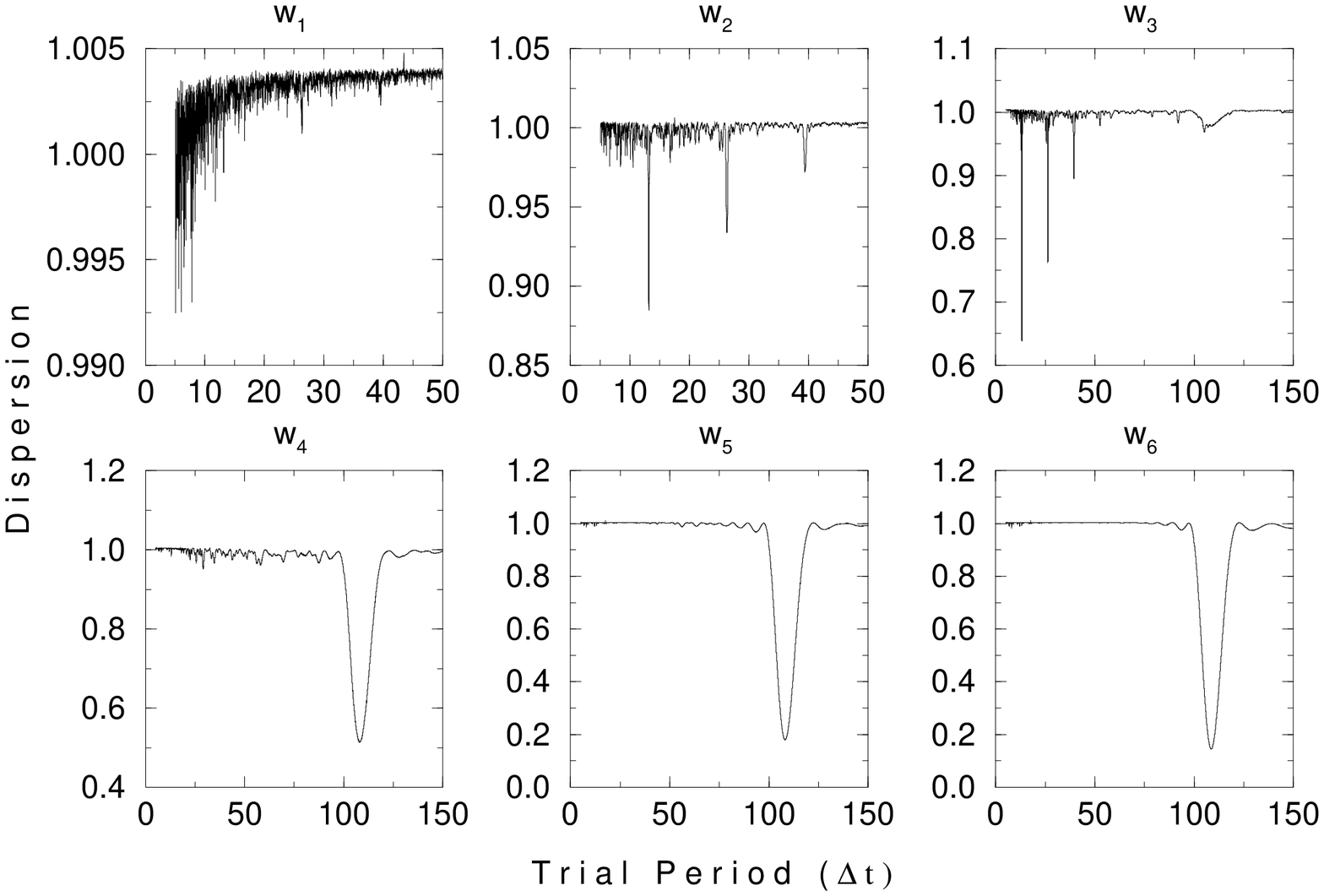}
\caption[]{{\sevensize WPDM} output for each wavelet plane of the data at the top of
Fig.~\ref{gauss_13_0.5}.}
\label{gauss_13_0.5.wpdm}
\end{figure}

\begin{figure}
\mbox{}
\vspace{6.0cm}
\includegraphics{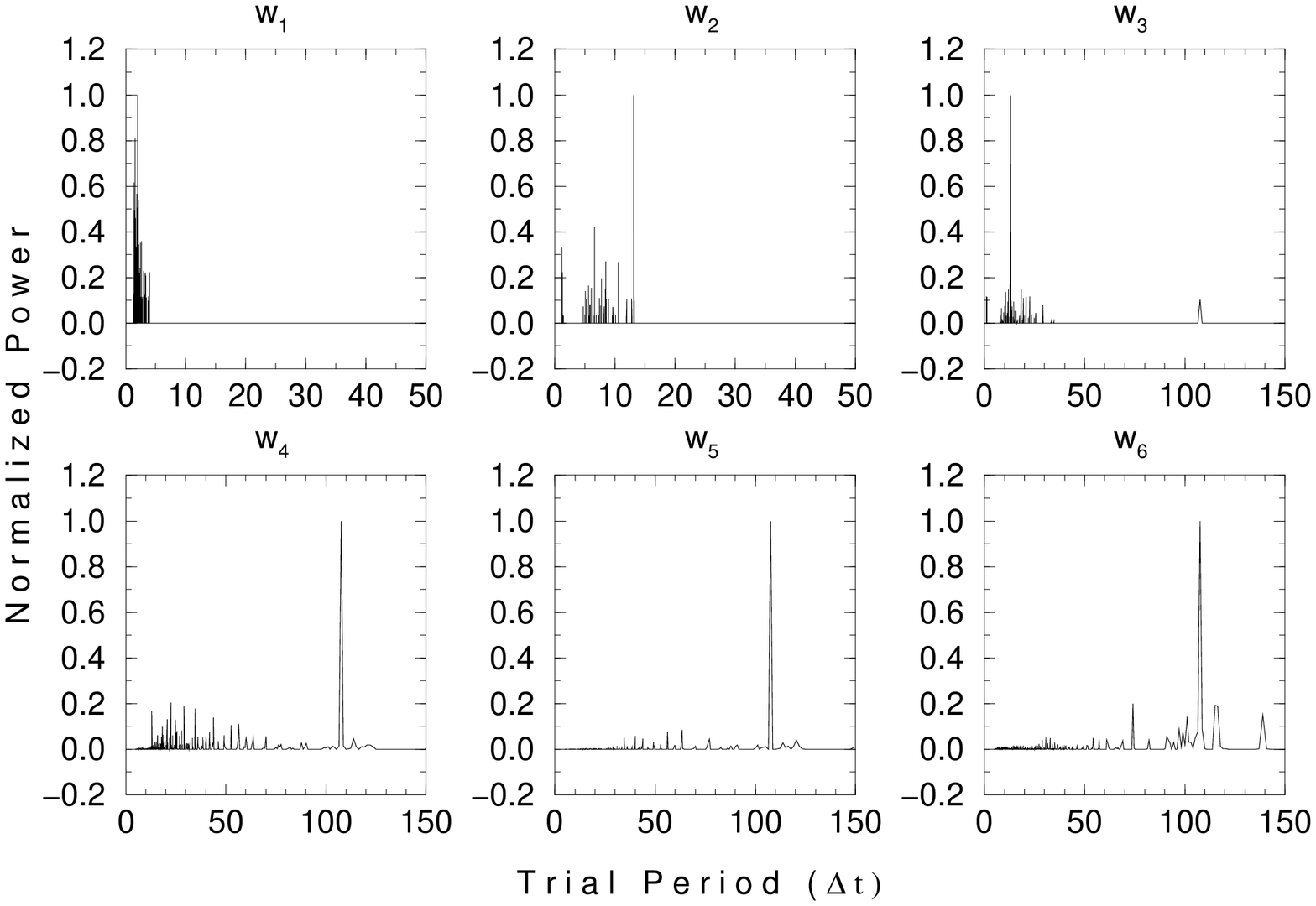}
\caption[]{{\sevensize WCLEAN} output for each wavelet plane of the data at the top of
Fig.~\ref{gauss_13_0.5}.}
\label{gauss_13_0.5.wclean}
\end{figure}

\begin{figure}
\mbox{}
\vspace{6.7cm}
\includegraphics{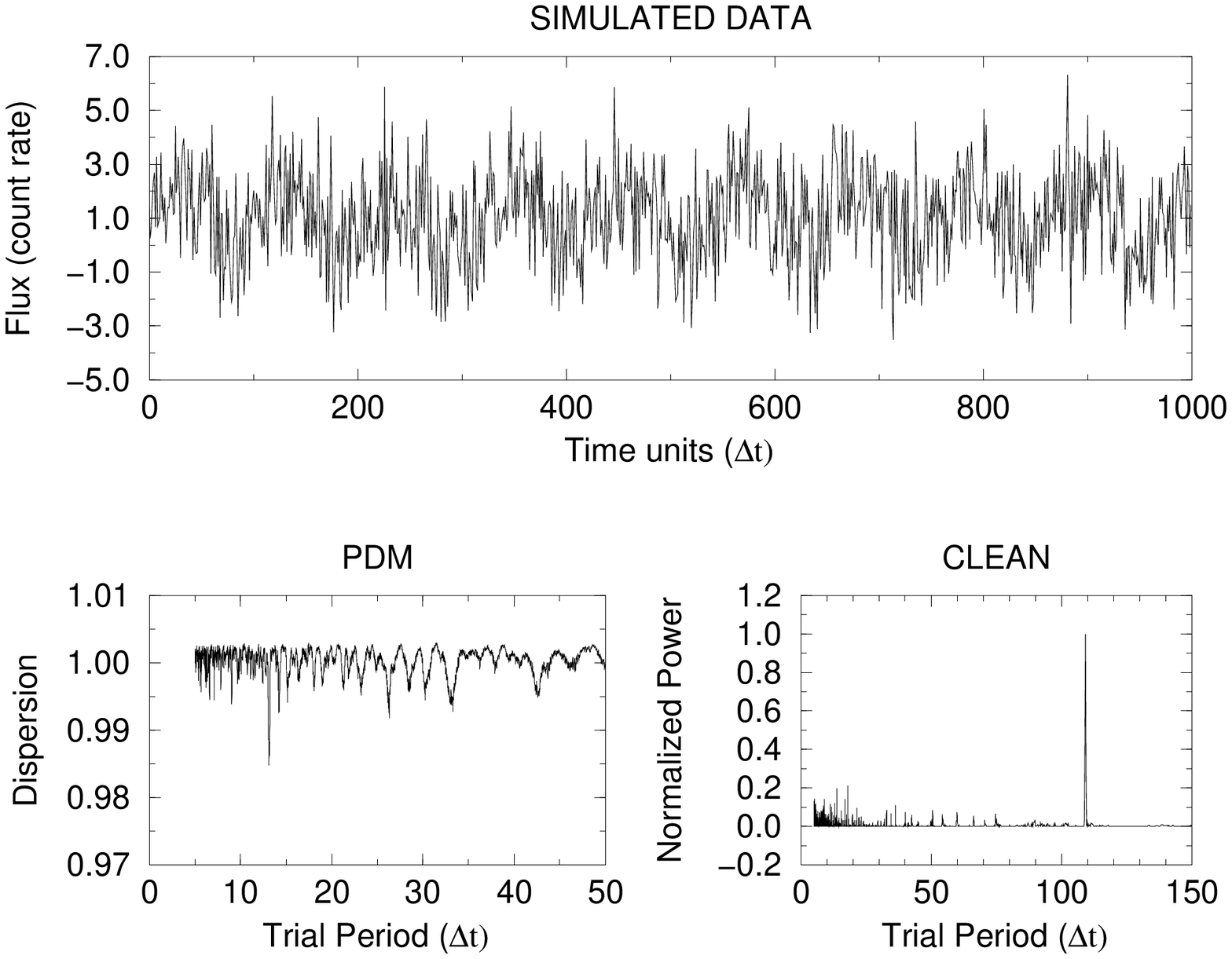}
\caption[]{Top: simulated data of a Gaussian with 13.13 units period, a FWHM of 6.0 units, amplitude=0.5 and $\sigma_{\mathrm noise}=1.5$, superimposed on the sinusoid with 108.5 units period and amplitude=1.0. Bottom: {\sevensize PDM} and {\sevensize CLEAN} outputs of this data set.}
\label{gauss_13_1.5}
\end{figure}

\begin{figure}
\mbox{}
\vspace{6.0cm}
\includegraphics{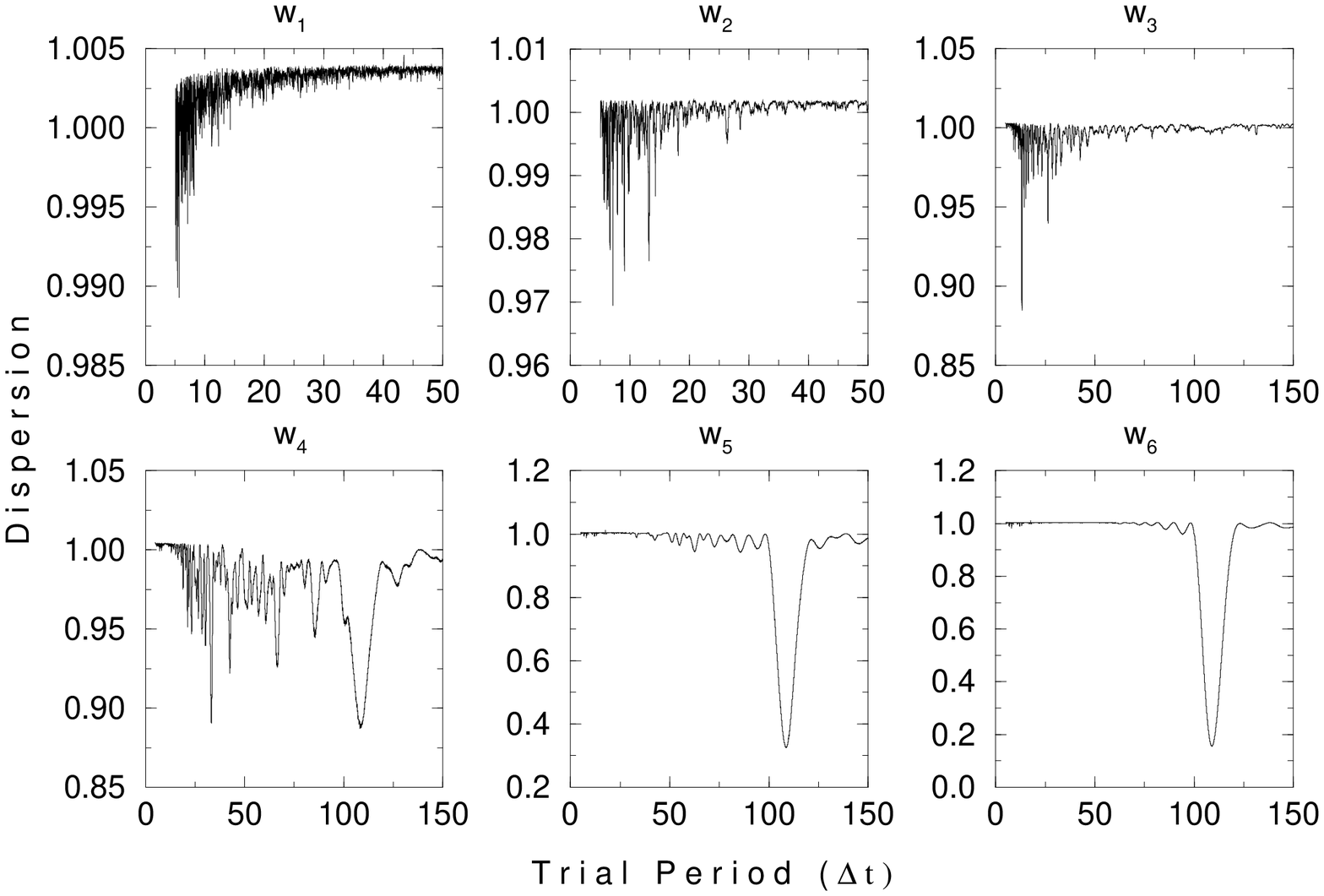}
\caption[]{{\sevensize WPDM} output for each wavelet plane of the data at the top of
Fig.~\ref{gauss_13_1.5}.}
\label{gauss_13_1.5.wpdm}
\end{figure}

\begin{figure}
\mbox{}
\vspace{6.0cm}
\includegraphics{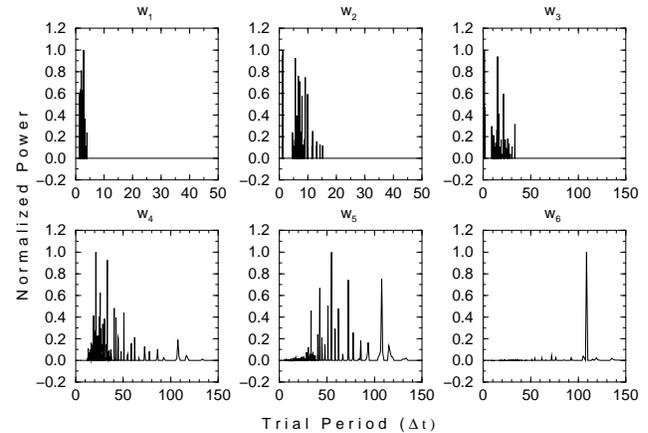}
\caption[]{{\sevensize WCLEAN} output for each wavelet plane of the data at the top of
Fig.~\ref{gauss_13_1.5}.}
\label{gauss_13_1.5.wclean}
\end{figure}

Finally, and for illustrative purposes, we show in Fig.~\ref{gauss_13_0.5} and
Fig.~\ref{gauss_13_1.5} two simulated data sets generated with the following
common parameters: 13.13 units period, FWHM=6.0 units and amplitude=0.5. The
only difference is that in Fig.~\ref{gauss_13_0.5}, $\sigma_{\mathrm
noise}=0.5$, while in Fig.~\ref{gauss_13_1.5}, $\sigma_{\mathrm noise}=1.5$.
The outputs of {\sevensize PDM}, {\sevensize CLEAN}, {\sevensize WPDM} and
{\sevensize WCLEAN} are shown from Fig.~\ref{gauss_13_0.5} to
Fig.~\ref{gauss_13_1.5.wclean}.

As can be seen in the figures, in the case $\sigma_{\mathrm noise}=0.5$, the
13.13 units period is clearly detected by all methods. In {\sevensize WPDM} and
{\sevensize WCLEAN} the period is detected in the wavelet planes $w_2$ and
$w_3$. These detections are much better than those of {\sevensize PDM} and
{\sevensize CLEAN}.

On the other hand, in the $\sigma_{\mathrm noise}=1.5$ case, the period is not
detected by {\sevensize PDM} nor by {\sevensize CLEAN}. {\sevensize WCLEAN}
also fails to detect it: the maximum in the wavelet plane $w_3$ is not
indicative of a period detection and, moreover, its position is in the 15.5
units trial period. The only clear detection with a subharmonic is in $w_3$ of
{\sevensize WPDM}.

\section{Conclusions}
\label{conclusions}

When dealing with two superimposed periodic signals, the performance of
classical methods, like {\sevensize PDM} and {\sevensize CLEAN}, depends on the
characteristics of these signals. These methods often fail to detect
simultaneously both periods, even with high or moderate signal-noise ratios.

With the wavelet based methods we present here, {\sevensize WPDM} and
{\sevensize WCLEAN}, it is possible to reach lower values of signal-noise
ratios and still detect both periodic signals. This means that the wavelet
based methods are less affected by noise than {\sevensize PDM} and {\sevensize
CLEAN}, and they allow us to detect periodic signals with noisier data.

Another advantage of the proposed methods is the simultaneous detection of
periods in several wavelet planes. If a period is marginally detected in more
than one wavelet plane, we are probably detecting a true periodic signal. This
allows us to improve the detection confidence.

These facts prove the major ability of {\sevensize WPDM} and {\sevensize
WCLEAN} over {\sevensize PDM} and {\sevensize CLEAN} to deal with noisy signals
and to detect superimposed periods even with low signal-noise ratios.

\section*{Acknowledgments}
We acknowledge partial support from DGI of the Ministerio de Ciencia y Tecnolog\'\i a (Spain) under grant AYA2001-3092.
This research has made use of facilities of CESCA and CEPBA, coordinated by C4 (Centre de Computaci\'o i Comunicacions de Catalunya).
During this work, M.~Rib\'o has been supported by a fellowship from CIRIT (Generalitat de Catalunya, ref. 1999~FI~00199).

\bsp

\label{lastpage}

\end{document}